\documentclass[11pt]{article}

\usepackage{amsmath,amsthm,amsfonts,array}
\allowdisplaybreaks

\DeclareMathOperator{\Hom}{Hom}
\DeclareMathOperator{\Com}{Com}

\DeclareMathOperator{\1}{id}
\newcommand{\NN}{\mathbb{N}}
\newcommand{\RR}{\mathbb{R}}
\newcommand{\EEnd}{\mathcal End}
\newcommand{\EE}{\mathcal E}

\renewcommand{\=}{:=}
\renewcommand{\t}{\otimes}

\newcommand{\m}{\overset{\circ}{\mu}}

\newcommand{\pp}{\hat{p}}
\newcommand{\ppp}{p_0}
\newcommand{\q}{\hat{q}}
\newcommand{\muu}{\hat{\mu}}
\newcommand{\qxi}{\hat{\xi}}
\newcommand{\Q}{\hat{Q}}
\newcommand{\PP}{\hat{P}}
\newcommand{\ee}{\varepsilon}
\renewcommand{\H}{\hat{H}}
\newcommand{\e}{\hat{\varepsilon}}
\newtheorem{thm}{Theorem}[section]

 \newtheorem{lemma}[thm]{Lemma}
 \newtheorem{cor}[thm]{Corollary}

 \newtheorem{con}[thm]{Conjecture}

\theoremstyle{definition}
 \newtheorem{defn}[thm]{Definition}

\theoremstyle{definition}

\theoremstyle{definition}
 \newtheorem{rem}[thm]{Remark}

\numberwithin{equation}{section}
\numberwithin{table}{section}
\begin{document}

\title{\bf\LARGE Operadic quantization}
\author{\Large E. Paal and J. Virkepu}
\date{}
\maketitle
\thispagestyle{empty}

\begin{abstract}
By a simple example, quantization of some 3-spaces is explained.
\end{abstract}

\section{Introduction and outline of the paper}

In Hamiltonian formalism, a mechanical system is described by the
canonical variables $q^i,p_i$ and their time evolution is prescribed
by the Hamiltonian equations
\begin{equation}
\label{ham}
\dfrac{dq^i}{dt}=\dfrac{\partial H}{\partial p_i}, \quad
\dfrac{dp_i}{dt}=-\dfrac{\partial H}{\partial q^i}
\end{equation}
By a Lax representation \cite{Lax68} of a mechanical system
one means such a pair $(L,M)$ of matrices (linear operators) $L,M$
that the above Hamiltonian system may be represented as the Lax
equation
\begin{equation}
\label{lax}
\dfrac{dL}{dt}= ML-LM
\end{equation}
Thus, from the algebraic point of view, mechanical systems may be
represented by linear operators, i.e by  linear maps $V\to V$  of a
vector space $V$. As a generalization of this one can pose the
following question \cite{Paal07}: how can the time evolution of the linear   operations (multiplications) $V^{\t n}\to V$ be described?

The algebraic operations (multiplications) can be seen as an example
of the \emph{operadic} variables \cite{Ger}. If an operadic system
depends on time one can speak about \emph{operadic dynamics}
\cite{Paal07}. The latter may be introduced by simple and natural
analogy with the Hamiltonian dynamics. In particular, the time
evolution of the operadic variables may be given by the operadic Lax
equation. In \cite{PV07,PV08,PV08-1}, the low-dimensional binary operadic Lax representations for the harmonic oscillator were constructed.
In \cite{PV08-2} it was shown how the operadic Lax representations are related to the conservation of energy.

In this paper, the operadic Lax representations for the harmonic oscillator are used to construct the quantum counterparts of some real three dimensional Lie algebras.  The Jacobi operators of these quantum algebras are studied in semiclassical approximation.

The paper is organized as follows.
In Sec. \ref{sec:operad} we recall the definition of the endomorphim operad and introduce the Gerstenhaber brackets as a main tool for constructing the operadic Lax pairs.
In Sec. \ref{sec:Lax} we describe the idea of operadic dynamics by using the operadic Lax equation.
This idea is then illustrated in Sec. \ref{sec:3d}  and Sec. \ref{sec:initial} by using the harmonic oscillator and the natural initial conditions.
In Sec. \ref{sec:3d_Lie} and Sec. \ref{sec:dyn} the operadic Lax representations are then used to construct the dynamical deformations of real three dimensional Lie algebras VII$_{a}$, III$_{a=1}$, VI$_{a\neq1}$ from the Bianchi classification.
The quantum counterparts of these Lie algebras are constructed in Sec.
\ref{sec:q_Lie}. The quantum multiplication is defined in a natural way in the state space of the quantum harmonic oscillator. It turns out that the Jacobi identity is violated in these quantum algebras and to inquire situation we use in Sec.
\ref{a1} and Sec. \ref{a2} the semiclassical approximation. In this approximation one can explicitly see how the quantum mechanical fundamental canonical commutation relations spoil the Jacobi identity on the quantum algebras.

\section{Endomorphism operad and Gerstenhaber brackets}
\label{sec:operad}

Let $K$ be a unital associative commutative ring, $V$ be a unital
$K$-module, and $\EE_V^n\= {\EEnd}_V^n\= \Hom(V^{\t n},V)$
($n\in\NN$). For an \emph{operation} $f\in\EE^n_V$, we refer to $n$
as the \emph{degree} of $f$ and often write (when it does not cause
confusion) $f$ instead of $\deg f$. For example, $(-1)^f\= (-1)^n$,
$\EE^f_V\=\EE^n_V$ and $\circ_f\= \circ_n$. Also, it is convenient
to use the \emph{reduced} degree $|f|\= n-1$. Throughout this paper,
we assume that $\t\= \t_K$.

\begin{defn}[endomorphism operad \cite{Ger}]
\label{HG}
For $f\t g\in\EE_V^f\t\EE_V^g$ define the
\emph{partial compositions}
\[
f\circ_i g\= (-1)^{i|g|}f\circ(\1_V^{\t i}\t g\t\1_V^{\t(|f|-i)})
\quad \in\EE^{f+|g|}_V,
         \quad 0\leq i\leq |f|
\]
The sequence $\EE_V\= \{\EE_V^n\}_{n\in\NN}$, equipped with the
partial compositions $\circ_i$, is called the \emph{endomorphism
operad} of $V$.
\end{defn}

\begin{defn}[total composition \cite{Ger}]
The \emph{total composition}
$\circ \:\EE^f_V\t\EE^g_V\to\EE^{f+|g|}_V$ is defined by
\[
f\circ g\= \sum_{i=0}^{|f|}f\circ_i g\quad \in \EE_V^{f+|g|}, \quad |\circ|=0
\]
The pair $\Com\EE_V\= \{\EE_V,\circ\}$ is called the \emph{composition
algebra} of $\EE_V$.
\end{defn}

\begin{defn}[Gerstenhaber brackets \cite{Ger}]
The  \emph{Gerstenhaber brackets} $[\cdot,\cdot]$ are defined in
$\Com\EE_V$ as a graded commutator by
\[
[f,g]\= f\circ g-(-1)^{|f||g|}g\circ f=-(-1)^{|f||g|}[g,f],\quad
|[\cdot,\cdot]|=0
\]
\end{defn}

The \emph{commutator algebra} of $\Com \EE_V$ is denoted as
$\Com^{-}\!\EE_V\= \{\EE_V,[\cdot,\cdot]\}$. One can prove (e.g
\cite{Ger}) that $\Com^-\!\EE_V$ is a \emph{graded Lie algebra}. The
Jacobi identity reads
\[
(-1)^{|f||h|}[f,[g,h]]+(-1)^{|g||f|}[g,[h,f]]+(-1)^{|h||g|}[h,[f,g]]=0
\]

\section{Operadic dynamics and Lax equation}
\label{sec:Lax}

Assume that $K\= \RR$ or $K\= \mathbb{C}$ and operations are
differentiable. Dynamics in operadic systems (operadic dynamics) may
be introduced by

\begin{defn}[operadic Lax pair \cite{Paal07}]
Allow a classical dynamical system to be described by the
Hamiltonian system \eqref{ham}. An \emph{operadic Lax pair} is a
pair $(\mu,M)$ of homogeneous operations $\mu,M\in\EE_V$, such that the
Hamiltonian system  (\ref{ham}) may be represented as the
\emph{operadic Lax equation}
\[
\frac{d\mu}{dt}=[M,\mu]\= M\circ\mu-(-1)^{|M||\mu|}\mu\circ M
\]
The pair $(L,M)$ is also called an \emph{operadic Lax representations} of/for Hamiltonian system \eqref{ham}.
Evidently, the degree constraints $|M|=|L|=0$ give rise to ordinary
Lax equation (\ref{lax}) \cite{Lax68}. In this paper we assume that $|M|=0$.
\end{defn}

The Hamiltonian of the harmonic oscillator (HO) is
\[
H(q,p)=\frac{1}{2}(p^2+\omega^2 q^2)
\]
Thus, the Hamiltonian system of HO reads
\begin{equation*}
%\label{eq:h-osc}
\frac{dq}{dt}=\frac{\partial H}{\partial p}=p,\quad
\frac{dp}{dt}=-\frac{\partial H}{\partial q}=-\omega^2q
\end{equation*}
If $\mu$ is a linear algebraic operation we can use the above
Hamilton equations to obtain
\[
\dfrac{d\mu}{dt} =\dfrac{\partial\mu}{\partial
q}\dfrac{dq}{dt}+\dfrac{\partial\mu}{\partial p}\dfrac{dp}{dt}
=p\dfrac{\partial\mu}{\partial
q}-\omega^2q\dfrac{\partial\mu}{\partial p}
 =[M,\mu]
\]
Therefore, we get the following linear partial differential equation
for $\mu(q,p)$:
\begin{equation}
\label{eq:diff}
p\dfrac{\partial\mu}{\partial
q}-\omega^2q\dfrac{\partial\mu}{\partial p}=[M,\mu]
\end{equation}
By integrating \eqref{eq:diff} one can get collections of operations called \cite{Paal07} the \emph{operadic} (Lax representations for/of) harmonic oscillator.

\section{3D binary anti-commutative operadic Lax representations for harmonic oscillator}
\label{sec:3d}

\begin{lemma}
\label{lemma:harmonic3} Matrices
\[
L\=\begin{pmatrix}
    p & \omega q & 0 \\
    \omega q & -p & 0 \\
    0 & 0 & 1 \\
  \end{pmatrix},\quad
M\=\frac{\omega}{2}
\begin{pmatrix}
    0 & -1 &0\\
1 & 0 & 0\\
0 & 0 & 0
  \end{pmatrix}
\]
give a 3-dimensional Lax representation for the harmonic oscillator.
\end{lemma}

\begin{defn}[quasi-canonical coordinates]
For the HO, define its  \emph{quasi-canonical coordinates} $Q$ and $P$ by
\begin{equation*}
\label{eq:def_A}
P^{2} - Q^{2}=2p,\quad
QP=\omega q
\end{equation*}
\end{defn}

\begin{rem}
Note that these constraints easily imply
\[
P^2+Q^2=2\sqrt{2H}
\]
\end{rem}

\begin{thm}[\cite{PV08-1}]
\label{thm:main}
Let $C_{\nu}\in\mathbb{R}$ ($\nu=1,\ldots,9$) be
arbitrary real--valued parameters, such that
\begin{equation*}
%\label{eq:cond}
C_2^2+C_3^2+C_5^2+C_6^2+C_7^2+C_8^2\neq0
\end{equation*}
Let $M$ be defined as in Lemma \ref{lemma:harmonic3} and
$\mu: V\otimes V\to V$ be a binary operation in a 3 dimensional real vector space $V$ with the coordinates
\begin{equation}\label{eq:theorem}
\begin{cases}
\mu_{11}^{1}=\mu_{22}^{1}=\mu_{33}^{1}=\mu_{11}^{2}=\mu_{22}^{2}=\mu_{33}^{2}=\mu_{11}^{3}=\mu_{22}^{3}=\mu_{33}^{3}=0\\
\mu_{23}^{1}=-\mu_{32}^{1}=C_2p-C_3\omega q-C_4\\
\mu_{13}^{2}=-\mu_{31}^{2}=C_2p-C_3\omega q+C_4\\
\mu_{31}^{1}=-\mu_{13}^{1}=C_2\omega q+C_3p-C_1\\
\mu_{23}^{2}=-\mu_{32}^{2}=C_2\omega q+C_3p+C_1\\
\mu_{12}^{1}=-\mu_{21}^{1}=C_5 P + C_6 Q\\
\mu_{12}^{2}=-\mu_{21}^{2}=C_5 Q - C_6 P\\
\mu_{13}^{3}=-\mu_{31}^{3}=C_7 P + C_8 Q\\
\mu_{23}^{3}=-\mu_{32}^{3}=C_7 Q - C_8 P\\
\mu_{12}^{3}=-\mu_{21}^{3}=C_9
\end{cases}
\end{equation}
Then $(\mu,M)$ is an operadic Lax pair for HO.
\end{thm}

\section{Initial conditions}
\label{sec:initial}

Specify the coefficients $C_{\nu}$ in Theorem \ref{thm:main} by the
initial conditions
\[
\left. \mu\right|_{t=0}=\m{}_,\quad
\left.p\right|_{t=0}
=p_0,\quad \left. q\right|_{t=0}=0
\]
Denoting $E\=H|_{t=0}$, the latter together with \eqref{eq:def_A} yield the initial
conditions for $Q$ and $P$:
\[
\begin{cases}
\left.\left(P^{2}+Q^{2}\right)\right|_{t=0}=2\sqrt{2E}\\
\left.\left(P^{2}-Q^{2}\right)\right|_{t=0}=2p_0\\
\left.PQ\right|_{t=0}=0
\end{cases}
\quad \Longleftrightarrow \quad
\begin{cases}
\ppp\!\!>0\\
\left.P^{2}\right|_{t=0}=2p_0\\
\left.Q\right|_{t=0}=0
\end{cases}
\vee\quad
\begin{cases}
\ppp<0\\
\left.P\right|_{t=0}=0\\
\left.Q^2\right|_{t=0}=-2p_0
\end{cases}
\]
In what follows assume that $p_0>0$ and $P|_{t=0}=\sqrt{2p_0}$. The other cases
can be treated similarly. Note that in this case $p_0=\sqrt{2E}$. From \eqref{eq:theorem} we get the following linear system:
\begin{equation}
\label{eq:constants}
\left\{
  \begin{array}{lll}
C_1=\frac{1}{2}\left(\overset{\circ}{\mu}{}_{23}^{2}-\overset{\circ}{\mu}{}_{31}^{1}\right),&
C_2=\frac{1}{2\ppp}\left(\overset{\circ}{\mu}{}_{13}^{2}+\overset{\circ}{\mu}{}_{23}^{1}\right),&
C_3=\frac{1}{2\ppp}\left(\overset{\circ}{\mu}{}_{23}^{2}+\overset{\circ}{\mu}{}_{31}^{1}\right)\vspace{1mm}\\
C_4=\frac{1}{2}\left(\overset{\circ}{\mu}{}_{13}^{2}-\overset{\circ}{\mu}{}_{23}^{1}\right),&
C_5=\frac{1}{\sqrt{2\ppp}}\overset{\circ}{\mu}{}_{12}^{1},&
C_6=-\frac{1}{\sqrt{2\ppp}}\overset{\circ}{\mu}{}_{12}^{2}\vspace{1mm}\\
C_7=\frac{1}{\sqrt{2\ppp}}\overset{\circ}{\mu}{}_{13}^{3},&
C_8=-\frac{1}{\sqrt{2\ppp}}\overset{\circ}{\mu}{}_{23}^{3},&
C_9=\overset{\circ}{\mu}{}_{12}^{3}
\end{array}
\right.
\end{equation}

\section{VII$_{a}$, III$_{a=1}$, VI$_{a\neq1}$}
\label{sec:3d_Lie}

We study only the algebras VII$_{a}$, III$_{a=1}$, VI$_{a\neq1}$ from the Bianchi classification of the real three dimensional Lie algebras \cite{Landau80}.
The structure equations of the 3-dimensional real Lie algebras can be presented
as follows:
\[
[e_1,e_2]=-\alpha e_2+n^{3}e_3,\quad
[e_2,e_3]=n^{1}e_1,\quad
[e_3,e_1]=n^{2}e_2+\alpha e_3
\]
The values of the parameters $\alpha,n^{1}, n^{2},n^{3}$  and the corresponding structure constants for II, VII$_{a}$, III$_{a=1}$, VI$_{a\neq1}$  are presented in Table \ref{table:Bianchi1}. Note that II is the real three dimensional Heisenberg algebra.
\begin{table}[ht]
\begin{center}
\begin{tabular}{|c||c||c|c|c||c|c|c|c|c|c|c|c|c|c|c|}\hline
Bianchi type & $\alpha$ & $n^{1}$ & $n^{2}$ & $n^{3}$ &
$\overset{\circ}{\mu}{}_{12}^{1}$ &
$\overset{\circ}{\mu}{}_{12}^{2}$ &
$\overset{\circ}{\mu}{}_{12}^{3}$ &
 $\overset{\circ}{\mu}{}_{23}^{1}$ & $\overset{\circ}{\mu}{}_{23}^{2}$ & $\overset{\circ}{\mu}{}_{23}^{3}$
  & $\overset{\circ}{\mu}{}_{31}^{1}$ & $\overset{\circ}{\mu}{}_{31}^{2}$ &
  $\overset{\circ}{\mu}{}_{31}^{3}$\\\hline\hline
II & 0 & $1$ & 0 & 0 & 0 & 0 & 0 & $1$ & 0 & 0 & 0 & 0 & 0
\\\hline
VII$_{a}$& $a$ & 0 & $1$ & $1$ & 0 & $-a$ & $1$ & 0 & 0 & 0 & 0 &
$1$ & $a$ \\\hline
III$_{a=1}$& 1 & 0 & $1$ & $-1$ & 0 & $-1$ & $-1$ & 0 & 0 & 0 & 0 &
$1$ & $1$
\\\hline
VI$_{a\neq 1}$& $a$ & 0 & $1$ & $-1$ & 0 & $-a$ & $-1$ & 0 & 0 & 0 &
0 & $1$ & $a$
\\\hline
\end{tabular}
\end{center}
\caption{II, VII$_{a}$, III$_{a=1}$, VI$_{a\neq1}$. Here $a>0$.}
\label{table:Bianchi1}
\end{table}

\section{VII$_{a}^{t}$, III$_{a=1}^{t}$, VI$_{a\neq1}^{t}$}
\label{sec:dyn}

By using the structure constants of the 3-dimensional Lie algebras
in the Bianchi classification, Theorem \ref{thm:main} and relations
\eqref{eq:constants} one can propose that evolution of VII$_{a}$, III$_{a=1}$, VI$_{a\neq1}$ can be prescribed \cite{PV08-2}
as given in Table \ref{table:Bianchi3}.

\begin{table}[ht]
\begin{center}\setlength\extrarowheight{4pt}
\begin{tabular}{|c||c|c|c|c|c|c|c|c|c|c|c|}\hline
Dynamical Bianchi type & $\mu_{12}^{1}$ & $\mu_{12}^{2}$ &
$\mu_{12}^{3}$ & $\mu_{23}^{1}$ & $\mu_{23}^{2}$ & $\mu_{23}^{3}$ &
$\mu_{31}^{1}$ & $\mu_{31}^{2}$ &  $\mu_{31}^{3}$
\\[1.5ex]\hline\hline
VII$^{t}_a$ & $\frac{aQ}{\sqrt{2p_0}}$ &
$\frac{-aP}{\sqrt{2p_0}}$ & $1$ & $\frac{p-p_0}{-2p_0}$ &
$\frac{\omega q}{-2p_0}$ & $\frac{-aQ}{\sqrt{2p_0}}$ &
$\frac{\omega q}{-2p_0}$ & $\frac{p+p_0}{2p_0}$ &
$\frac{aP}{\sqrt{2p_0}}$
\\ [1.5ex] \hline
III$_{a=1}^{t}$ & $\frac{Q}{\sqrt{2p_0}}$ &
$\frac{-P}{\sqrt{2p_0}}$ & $-1$ & $\frac{p-p_0}{-2p_0}$ &
$\frac{\omega q}{-2p_0}$ & $\frac{-Q}{\sqrt{2p_0}}$ & $\frac{\omega
q}{-2p_0}$ & $\frac{p+p_0}{2p_0}$ & $\frac{P}{\sqrt{2p_0}}$
\\ [1.5ex] \hline
VI$_{a\neq1}^{t}$ & $\frac{aQ}{\sqrt{2p_0}}$ &
$\frac{-aP}{\sqrt{2p_0}}$ & $-1$ & $\frac{p-\ppp}{-2p_0}$ &
$\frac{\omega q}{-2p_0}$ & $\frac{-aQ}{\sqrt{2p_0}}$ &
$\frac{\omega q}{-2p_0}$ & $\frac{p+p_0}{2p_0}$ &
$\frac{aP}{\sqrt{2p_0}}$
\\ [1.5ex] \hline
\end{tabular}
\end{center}
\caption{VII$_{a}^{t}$, III$_{a=1}^{t}$, VI$_{a\neq1}^{t}$. Here $p_0=\sqrt{2E}$.}
\label{table:Bianchi3}
\end{table}

\section{VII$_{a}^{\hbar}$, III$_{a=1}^{\hbar}$, VI$_{a\neq1}^{\hbar}$ and quantum Jacobi operators}
\label{sec:q_Lie}

By using the algebras VII$_{a}^{t}$, III$_{a=1}^{t}$, VI$_{a\neq1}^{t}$
from Table \ref{table:Bianchi3}, one can propose \cite{PV09-1} their quantum counterparts
VII$_{a}^{\hbar}$, III$_{a=1}^{\hbar}$, VI$_{a\neq1}^{\hbar}$ as follows.

Let $\{e_1,e_2,e_3\}$ be the basis of  VII$_{a}$, III$_{a=1}$, VI$_{a\neq1}$ used in the Bianchi classification.
By using  Table \ref{table:Bianchi4} we define the structure equations of the quantum counterparts of VII$_{a}$, III$_{a=1}$, VI$_{a\neq1}$ by formal linear combinations
\[
[e_i,e_j]_\hbar:=\muu_{ij}^{s} e_s
\]
\begin{table}[ht]
\begin{center}\setlength\extrarowheight{4pt}
\begin{tabular}{|c||c|c|c|c|c|c|c|c|c|c|c|}\hline
Quantum Bianchi type & $\muu_{12}^{1}$ & $\muu_{12}^{2}$ &
$\muu_{12}^{3}$ & $\muu_{23}^{1}$ & $\muu_{23}^{2}$ &
$\muu_{23}^{3}$ & $\muu_{31}^{1}$ & $\muu_{31}^{2}$ &
$\muu_{31}^{3}$
\\[1.5ex]\hline\hline
VII$^{\hbar}_a$ & $\frac{a\Q}{\sqrt{2\ppp}}$ &
$\frac{-a\PP}{\sqrt{2\ppp}}$ & $1$ & $\frac{\pp-\ppp}{-2\ppp}$ &
$\frac{\omega \q}{-2\ppp}$ & $\frac{-a\Q}{\sqrt{2\ppp}}$ &
$\frac{\omega \q}{-2\ppp}$ & $\frac{\pp+\ppp}{2\ppp}$ &
$\frac{a\PP}{\sqrt{2\ppp}}$
\\ [1.5ex] \hline
III$_{a=1}^{\hbar}$ & $\frac{\Q}{\sqrt{2\ppp}}$ &
$\frac{-\PP}{\sqrt{2\ppp}}$ & $-1$ & $\frac{\pp-\ppp}{-2\ppp}$ &
$\frac{\omega \q}{-2\ppp}$ & $\frac{-\Q}{\sqrt{2\ppp}}$ & $\frac{\omega
\q}{-2\ppp}$ & $\frac{\pp+\ppp}{2\ppp}$ & $\frac{\PP}{\sqrt{2\ppp}}$
\\ [1.5ex] \hline
VI$_{a\neq1}^{\hbar}$ & $\frac{a\Q}{\sqrt{2\ppp}}$ &
$\frac{-a\PP}{\sqrt{2\ppp}}$ & $-1$ & $\frac{\pp-\ppp}{-2\ppp}$ &
$\frac{\omega \q}{-2\ppp}$ & $\frac{-a\Q}{\sqrt{2\ppp}}$ &
$\frac{\omega \q}{-2\ppp}$ & $\frac{\pp+\ppp}{2\ppp}$ &
$\frac{a\PP}{\sqrt{2\ppp}}$
\\ [1.5ex] \hline
\end{tabular}
\end{center}
\caption{VII$_{a}^{\hbar}$, III$_{a=1}^{\hbar}$, VI$_{a\neq1}^{\hbar}$.}
\label{table:Bianchi4}
\end{table}
For $x,y\in {\rm VII}_{a}, {\rm III}_{a=1}, {\rm VI}_{a\neq1}$, their quantum multiplication is defined by
\[
[x,y]_\hbar
:=\hat{\mu}^{i}_{jk} x^{j}y^{k} e_i
=\hat{\mu}^{1}_{jk} x^{j}y^{k} e_1
+\hat{\mu}^{2}_{jk} x^{j}y^{k} e_2
+\hat{\mu}^{3}_{jk} x^{j}y^{k} e_3
\]
Then the quantum Jacobi operator is defined by
\begin{align*}
\hat{J}_\hbar(x;y;z)
&:=[x,[y,z]_\hbar]_\hbar+[y,[z,x]_\hbar]_\hbar+[z,[x,y]_\hbar]_\hbar\\
&\,\,=\hat{J}^1_\hbar(x;y;z)e_1+\hat{J}^2_\hbar(x;y;z)e_2+\hat{J}^3_\hbar(x;y;z)e_3
\end{align*}
 In \cite{PV09-1, PV10-1} the quantum Jacobi operators were calculated for all real three dimensional Lie algebras. Here we concentrate only on
 VII$^{\,\hbar}_a$, III$_{a=1}^{\,\hbar}$, VI$_{a\neq1}^{\,\hbar}$.
Denote
\begin{equation*}
(x,y,z)\=
\begin{vmatrix}
 x^{1} & x^{2} & x^{3} \\
 y^{1} & y^{2} & y^{3} \\
z^{1} & z^{2} & z^{3} \\
\end{vmatrix},\quad
\qxi^{1}\=\omega\q\Q + (\pp - \ppp)\PP, \quad
\qxi^{2}\=\omega\q\PP - (\pp + \ppp)\Q
\end{equation*}
Recall
\begin{thm}[\cite{PV09-1, PV10-1}]
The Jacobi operator components of VII$^{\,\hbar}_a$ and
VI$_{a\neq1}^{\,\hbar}$ read
\[
\hat{J}^{1}_\hbar(x;y;z)=-\frac{a(x,y,z)}{\sqrt{2\ppp^{3}}}\qxi^{1},\quad
\hat{J}^{2}_\hbar(x;y;z)=-\frac{a(x,y,z)}{\sqrt{2\ppp^{3}}}\qxi^{2},\quad
\hat{J}^{3}_\hbar(x;y;z)=\frac{a^{2}(x,y,z)}{\ppp}[\PP,\Q]
 \]
For III$_{a=1}^{\,\hbar}$ one has the same formulae with $a=1$.
\end{thm}

\section{Semiclassical quantum conditions}
\label{a1}

\begin{thm}[Poisson brackets of quasi-canonical coordinates]
\label{eq:quasi_poisson}
The quasi-canonical coordinates $Q$ and $P$ satisfy the relations
\begin{equation*}
%\label{eq:poisson_A}
\{P,P\}=0=\{Q,Q\},\quad \{P,Q\}=\ee\=\frac{\omega}{2\sqrt{2H}}
\end{equation*}
\end{thm}

\begin{proof}
Calculate:
\begin{align*}
2\omega=2\omega \{p,q\}
&=\{P^2-Q^2, PQ\}\\
&=\{P^2, PQ\}-\{Q^2, PQ\}\\
&=P\{P^2, Q\}-\{Q^2,P\}Q\\
&=P\{PP, Q\}-\{QQ,P\}Q\\
&=2(P^2+Q^2)\{P,Q\}\\
&=4\sqrt{2H}\{P,Q\}
\tag*{\qed}
\end{align*}
\renewcommand{\qed}{}
\end{proof}

When performing the quantization of the quasi-canonical variables,
 we shall use the Schr\"o\-dinger picture, i.e the operators $\q,\pp,\H$ and $\Q,\PP$ do not depend on time.
Denote by $[\cdot,\cdot]$ the ordinary commutator bracketing.
Following the canonical quantization prescription, the quasi-canonical coordinates would satisfy in the semiclassical limit ($\hbar\to 0$) the constraints
\begin{equation}
\label{eq:def_qA}
\PP^2+\Q^2\approx 2\sqrt{2\H},\quad
\PP^2-\Q^2\approx 2\pp,\quad
\PP\Q+\Q\PP\approx  2\omega \q
\end{equation}
and the \emph{quasi-canonical commutation relations} (quasi-CCR)  read as follows:
\begin{equation}
\label{eq:qpoisson_A}
[\PP,\PP]=0=[\Q,\Q],\quad
[\PP,\Q]=\frac{\hbar}{i}\e \=\frac{\hbar}{i}\frac{\omega}{2\sqrt{2\hat{H}}}
\end{equation}

\section{Semiclassical approximation of the Jacobi operator}
\label{a2}

\begin{thm}
We have
\begin{align*}
\hat{J}^{1}_\hbar(x;y;z)
&\approx \frac{a(x,y,z)}{\sqrt{2p_0^3}}
\left[\PP\left(\sqrt{2E} - \sqrt{2\H}\right)
 - \frac{\hbar}{i}\Q \frac{\e}{2}  \right]\\
\hat{J}^{2}_\hbar(x;y;z)
&\approx  \frac{a(x,y,z)}{\sqrt{2p_0^3}}
\left[\Q\left(\sqrt{2E} - \sqrt{2\H}\right)
+ \frac{\hbar}{i}\PP \frac{\e}{2} \right]\\
\hat{J}^{3}_\hbar(x;y;z)
&\approx \frac{\hbar}{i}  \frac{a^{2}(x,y,z)}{p_0} \e
\tag*{\qed}
\end{align*}
\end{thm}

\begin{proof}
Using relations \eqref{eq:def_qA} and \eqref{eq:qpoisson_A} first calculate:
\begin{align*}
\qxi^{1}
&\= \omega\q\Q + (\pp - p_0)\PP\\
&\approx \frac{1}{2}(\PP\Q+\Q\PP)\Q +\frac{1}{2}(\PP^{2}-\Q^{2})\PP - p_0 \PP\\
&\approx \frac{1}{2}(\PP\Q^{2}+\Q\PP\Q +\PP^{3} - \Q^{2}\PP)  -p_0\PP\\
&\approx \frac{1}{2}\left[\Q(\PP\Q - \Q\PP)+\PP(\Q^{2} + \PP^{2})\right]-p_0\PP\\
&\approx \frac{1}{2}\Q[\PP,\Q]+\frac{1}{2}\PP(\PP^{2} + \PP^{2})  -p_0\PP\\
&\approx \frac{\hbar}{i}\Q\frac{\e}{2} +\PP\sqrt{2\H} - \sqrt{2E}\PP\\
&\approx \frac{\hbar}{i}\Q\frac{\e}{2} +\PP\left(\sqrt{2\H} - \sqrt{2E}\right)
 \end{align*}
Next calculate
\begin{align*}
\qxi^{2}
&\= \omega\q\PP- (\pp + p_0)\Q\\
&\approx \frac{1}{2}(\PP\Q+\Q\PP)\PP-\frac{1}{2}(\PP^{2}-\Q^{2})\Q -p_0 \Q\\
&\approx \frac{1}{2}(\PP\Q\PP+\Q\PP^{2} -\PP^{2}\Q+ \Q^{3})  -p_0\Q\\
&\approx \frac{1}{2}\left[\PP(\Q\PP - \PP\Q)+\Q(\PP^{2} + \Q^{2})\right]-p_0\Q\\
&\approx -\frac{1}{2}\PP[\PP,\Q]+\frac{1}{2}\Q(\PP^{2} + \Q^{2})  -p_0\Q\\
&\approx -\frac{\hbar}{i}\PP\frac{\e}{2} +\Q\sqrt{2\H} - \sqrt{2E} \Q\\
&\approx -\frac{\hbar}{i}\PP\frac{\e}{2} +\Q\left(\sqrt{2\H} - \sqrt{2E}\right)
\tag*{\qed}
\end{align*}
\renewcommand{\qed}{}
\end{proof}

\begin{cor}
\label{H=E}
Let  the energy conservation law $\H=E$ hold. Then we have
\begin{align*}
\hat{J}^{1}_\hbar(x;y;z)
&\approx
-\frac{\hbar}{i} \frac{a(x,y,z)}{\sqrt{(2p_0)^3}} \frac{\omega}{2\sqrt{2E}} \Q
\\
\hat{J}^{2}_\hbar(x;y;z)
&\approx
\frac{\hbar}{i} \frac{a(x,y,z)}{\sqrt{(2p_0)^3}} \frac{\omega}{2\sqrt{2E}} \PP
\\
\hat{J}^{3}_\hbar(x;y;z)
&\approx
\frac{\hbar}{i}  \frac{a^{2}(x,y,z)}{p_0} \frac{\omega}{2\sqrt{2E}}
\end{align*}
\end{cor}

\section{Interpretation}

\begin{thm}
In semiclassical approximation, the Jacobi operator components generate a 3-dimensional real anti-commutative algebra with structure equations
\begin{equation}
\label{eq:TA}
[\hat{J}^{1}_\hbar,\hat{J}^{3}_\hbar]
\approx 0
\approx [\hat{J}^{2}_\hbar,\hat{J}^{3}_\hbar], \quad
[\hat{J}^{1}_\hbar,\hat{J}^{2}_\hbar]\approx C\hat{J}^{3}_\hbar
\end{equation}
where
\[
C\=-\left(\frac{\hbar\omega}{2E}\right)^{2}\frac{(x,y,z)}{32}
\]
\end{thm}

\begin{proof}
Use Corollary \ref{H=E}.
\end{proof}

\begin{defn}[derivative algebra]
The anti-commutative algebra given by the structure equations \eqref{eq:TA} is called the \emph{derivative algebra} of  VII$^{\,\hbar}_a$,
III$_{a=1}^{\,\hbar}$, VI$_{a\neq1}^{\,\hbar}$.
\end{defn}

\begin{cor}
Define the new basis in the derivative algebra:
\[
e_1=-(x,y,z)\hat{J}^{3}_\hbar, \quad
e_2=-(x,y,z)\hat{J}^{1}_\hbar, \quad
e_3=-(x,y,z)\hat{J}^{2}_\hbar
\]
Then
\[
\quad [e_1,e_3]\approx 0\approx [e_2,e_3], \quad [e_2,e_3]\approx \beta^{2}e_1\]
where
\[
\beta\= \frac{\hbar\omega}{2E} \frac{|(x,y,z)|}{4\sqrt{2}}
\]
\end{cor}

\begin{proof}
Calculate:
\begin{align*}
[e_2,e_3]
=(x,y,z)(x,y,z)[\hat{J}^{1}_\hbar,\hat{J}^{2}_\hbar]
\approx C(x,y,z)(x,y,z)\hat{J}^{3}_\hbar
=-C(x,y,z)e_1
=\beta^{2}e_1
\tag*{\qed}
\end{align*}
\renewcommand{\qed}{}
\end{proof}

\begin{con}
The derivative algebra of
VII$^{\,\hbar}_a$, III$_{a=1}^{\,\hbar}$, VI$_{a\neq1}^{\,\hbar}$   is the 3-dimensional real Heisenberg algebra.
\end{con}

\begin{proof}[Idea of proof]
By elementary calculus one can see that the Jacobi identity holds.
As the only non-vanishing structure constant is $\overset{\circ}{\mu}{}_{23}^{1}$, one can easily see from the Bianchi Table \ref{table:Bianchi1} that
$\beta\approx 1$ perfectly suits.
\end{proof}

\begin{cor}
In VII$^{\,\hbar}_a$, III$_{a=1}^{\,\hbar}$, VI$_{a\neq1}^{\,\hbar}$   we have
$|(x,y,z)|\approx 4\sqrt{2}(2n+1)$, $n=0,1,2,\dots$
\end{cor}

\begin{proof}
Use $\beta\approx1$ and $E=\hbar\omega(n+1/2)$.
\end{proof}

\begin{rem}
One may also consider the option $\beta:=L^{3}_{Planck}/2\sqrt{2}$.
\end{rem}

\section*{Acknowledgements}

The research was in part supported by the Estonian Science Foundation, Grant ETF-6912. The authors are grateful to A.~Fialowski for help concerning the Bianchi classification and to S.~Hervik and P.~Kuusk for help about using the Bianchi classification in cosmology.

\noindent
Tallinn University of Technology,
Ehitajate tee 5, 19086 Tallinn, Estonia

\end{document}